\documentclass[prb,showpacs,twocolumn]{revtex4}
\usepackage{psfig}
\begin{document}
 
\title{
Magnetic phenomena in 5d transition metal nanowires
}

\author{A. Delin}
\affiliation{Abdus Salam International Center for Theoretical Physics (ICTP), Strada Costiera 11, 34100 Trieste, Italy}
\author{E. Tosatti}
\affiliation{Abdus Salam International Center for Theoretical Physics (ICTP), Strada Costiera 11, 34100 Trieste, Italy\\
             International School for Advanced Studies (SISSA), via Beirut 2--4, 34014 Trieste, Italy\\
	     INFM DEMOCRITOS National Simulation Center, via Beirut 2--4, 34014 Trieste, Italy
	     }

\date{\today}
\begin{abstract}
We have carried out fully relativistic full-potential, spin-polarized, all-electron
density-functional calculations for straight, monatomic nanowires of 
the $5d$ transition and noble metals
Os, Ir, Pt and Au.
We find that, of these metal nanowires, Os and Pt have mean-field magnetic moments for values of the 
bond length at equilibrium. In the case of Au and Ir, the wires need to be slightly stretched in order to spin polarize.
An analysis of the band structures of the wires indicate that 
the superparamagnetic state that our calculations suggest will affect the
conductance through the wires --- though not by a large amount --- at least in the
absence of magnetic domain walls. 
It should thus lead to a characteristic temperature- and field dependent conductance,
and may also cause a significant spin polarization of the transmitted current.
\end{abstract}

\maketitle

\section{introduction}
There is presently a strong interest in the physics of metal nanowires of atomic dimensions. 
A freely hanging metallic nanowire is formed when two pieces of material, initially at contact, are pulled away from 
each other over atomic distances.
In the process a connective bridge or neck elongates and narrows.
Experimentally, segments of such nanowires have been formed between tips,
in particular of Au,\cite{kondo2000_helical,rodrigues2000} 
but very recently also in break junctions of Pt and Ir.\cite{smit2001}. 
The one-dimensional character of nanowires cause several
new physical phenomena to appear, like quantized conductance\cite{wees1988}
and helical geometries.\cite{gulseren1998, kondo2000_helical,tosatti2001_tension}
With respect to the bulk metal, the freely hanging nanowires are of course 
unstable and therefore transient objects, which undergo thinning and
eventual breaking. Nevertheless, the kinetics leading to that thinning 
process slows down when reaching special ``magic'' geometries, where 
long lifetimes of several seconds have been recorded. The occurrence
of magic geometries has been proposed to correspond to local minima of
the nanowire effective string tension.\cite{tosatti2001_tension}
Some of these wires are only a few atomic layers thick, but can extend in length
up to 15 nm, which corresponds to roughly 50 gold-atom diameters.\cite{kondo2000_helical}
Ultimately thin wires, consisting of only a single atomic strand, so-called monowires --- necessarily much shorter than
the above mentioned thicker wires --- have also been observed.\cite{smit2001,ohnishi1998,yanson1998}
Besides these transient objects, 
another type of nanowires, which are stable, also exist. 
Structurally stable nanowires can be grown on stepped surfaces, like for example the recently observed
Co monatomic chains on Pt substrate,\cite{gambardella2002} or inside tubular structures, like
the Ag nanowires of micrometer lengths grown inside self-assembled calix[4]hydroquinine nanotubes.\cite{heehong2001} 

An interesting question is of course if, when and how magnetism may appear in
nanowires and how --- if that is the case --- this affects the other properties.
Those metals, which are magnetic already in bulk, can be expected to be magnetic also as nanowires, Hund's rule being
reinforced by lower coordination.
But may also normally non-magnetic metals magnetize in nanowires form?
It has been suggested that even a jellium confined in a thin cylinder may in principle 
magnetize for certain radii of the cylinder.\cite{zabala1998}
However, the moment formation is confined to very special radii or electron densities,
and the associated energy gain is very small. 
That is of course so because exchange interactions, as described by Hund's first rule, are not particularly
strong in an $sp$ band metal (e.g., Na or Al), a typical system that might be thought of as a jellium.
The situation is radically different for transition metals of the $4d$ and $5d$
series. Because of the partly occupied $d$ orbitals, their ability to magnetize is 
much stronger and of a fundamentally different nature compared to the jellium.
In bulk, the resulting large exchange interactions of these metals are overwhelmed by the large
electron kinetic energies, resulting in very large bandwidths and a nonmagnetic ground state. 

In the present work, we concentrate on monatomic wires made up of the $5d$ elements
Os, Ir, Pt and Au, investigating the  possibility of
ferromagnetism\cite{note1} in these nanowires.
Of these metals, Os and Ir exhibit monolayer magnetism on Au or Ag 
substrates.\cite{blugel1992,redinger1995}
Os, Ir, and Pt might conceivably
develop Hund's rule magnetism in free-hanging nanowire form, due to their partly empty $d$ shell. 
Au, on the other hand, is basically an $sp$ metals
but with some $d$ orbitals quite close to the Fermi level, making it a borderline case.
If strong Hund's rule magnetism develops in these wires, the number of conductance channels would be greatly 
influenced, and the results could show up in the form of, e.g., strong 
and unusual joint magnetic field and temperature dependence in the
ballistic conductance. 

Of course, thermal fluctuations in nanowires are expected to be very large, 
which would destroy long range magnetic order in the absence of an external 
magnetic field.  Depending on temperature and on external field, there 
will nevertheless be two different fluctuation regimes: a slow one and a
fast one. 

Slow fluctuations 
such as those attainable at low temperatures and/or in presence of a
sufficiently large external field take a nanomagnet to a superparamagnetic
state, where magnetization fluctuates between equivalent magnetic 
valleys, separated by, e.g., anisotropy-induced energy
barriers. If the barriers are sufficiently large
the nanosystem spends most of the time within a single magnetic valley, 
and will for many practical purposes behave as magnetic. We may under 
these circumstances be allowed to neglect fluctuations altogether, and to
approximate some properties of the superparamagnetic nanosystem 
with those of a statically magnetized one. Experimentally, evidence of 
one-dimensional superparamagnetism  with fluctuations sufficiently slow on the time 
scale of the probe was recently reported for Co atomic chains deposited at Pt 
surface steps.\cite{gambardella2002}

At the opposite extreme --- a situation reached for example
at high temperatures, and in zero external field --- the energy barriers are 
so readily overcome that the magnetic state will be totally washed away
by fast fluctuations, leading to a conventional paramagnetic state.
A complete description of this high entropy state is beyond scope here, 
and we have chosen, as is usually done, to approximate it with the 
conventional $T=0$ nonmagnetic, singlet solution of the Kohn-Sham
electronic structure equations.

In this paper we will only deal with straight undimerized wires. This might appear 
oversimplified, since, for instance, it has been calculated that infinite 
gold wires have a local energy minimum for a zigzag structure.\cite{sanchezportal1999}
Similar structures are also possible for the other metals studied here. 
Our rationale for this simplification of the wire geometry is that wires
extended between two tips are inevitably subject to stretching. The
simple thermodynamics causing the wire-tip atom flow of atoms and
driving the thinning\cite{torres1999} implies a finite string
tension.\cite{tosatti2001_tension} Thus, even if a free-ended wire
favored a zigzag structure, this effect will be washed out in the ultimate 
wire hanging between tips just before breaking of the contact.
Moreover possible dimerization of the wires, an issue whose possible
relevance is restricted to gold, will not be considered here. 

\section{method}
In the present density-functional-based\cite{dft} electronic-structure calculations
we used the all-electron full-potential linear
muffin-tin orbital method (FP-LMTO).\cite{wills} 
This method assumes no shape approximation of the potential or wave functions.
The calculations were performed using the generalized gradient approximation
(GGA).\cite{gga} 
As a test, some calculations were also performed using the local density
approximation (LDA),\cite{lda}, giving results very similar to the GGA ones. 
Further, some calculations were double-checked using the WIEN code,\cite{wien97} again with 
very similar results.
We chose an all-electron approach in order to rid our calculations of 
possible sources of doubt that
may arise when using pseudopotentials in the presence of magnetism 
and in non-standard configurations.

The calculations were performed with inherently three-dimensional
codes, and thus the system simulated was an infinite two-dimensional
array of infinitely long, straight wires.
A one-dimensional Brillouin zone was used, i.e 
the k-points form a single line, stretching along the 
$z$-axis of the wire. 
The Bravais lattice in the $xy$-plane was chosen hexagonal. 
Furthermore, we used non-overlapping muffin-tin spheres with a constant radius in the calculations
of the equilibrium bond lengths $d$. 
The magnetic moments, bands structures, conductance-channel curves and  band widths were calculated using 
muffin-tin spheres scaling with the bond length. 
Convergence of the magnetic moment was ensured with respect to k-point mesh density, Fourier mesh density,
tail energies, and wire-wire vacuum distance.

We performed both scalar relativistic (SR) calculations, and calculations including the spin-orbit
coupling as well as the scalar-relativistic terms. 
The latter will be referred to as ``fully relativistic'' (FR) calculations in the following,
although we are not strictly solving the full Dirac equation, or making use of current density functional theory.
In the FR calculations, the spin axis was chosen to be 
aligned along the wire direction.

\section{results and discussion}

\begin{table}
\caption{
\label{tab:lattpar}
Calculated monowire and bulk equilibrium bond length $d$.  
Also shown are the calculated magnetic moments per atom, with and without spin-orbit coupling,
at the monowire equilibrium bond length. 
The right-most column displays the experimental ground-state
configuration for the free atoms. 
}
\begin{ruledtabular}
\begin{tabular}{cccccccl}
& wire            & wire         & bulk         &  bulk         & moment        & moment      & free         \\
&  $d$ ({\AA})    &  $d$ ({\AA}) & $d$ ({\AA})  &  $d$ ({\AA})  & ($\mu_B$)     & ($\mu_B$)   & atom         \\
metal & SR              & FR           & FR           &  exp.         & SR            & FR          & moment       \\
\hline
\hline
&      &      &       &       &      &      &                         \\     
Os & 2.31 & 2.30 & 2.76  & 2.73  & 1.3  & 0.3  &  4 ($^5 D_{4}$)          \\  
Ir & 2.31 & 2.34 & 2.75  & 2.71  & 0.8  & -    &  3 ($^4 F_{9/2}$)        \\  
Pt & 2.42 & 2.48 & 2.79  & 2.75  & -    & 0.6  &  2 ($^3 D_{3}$)          \\  
Au & 2.66 & 2.61 & 2.90  & 2.88  & -    & -    &  1 ($^2 S_{1/2}$)        \\ 
\end{tabular}

\end{ruledtabular}
\end{table}

\subsection{Bond lengths and energetics}
The chemical bonding in a wire is, of course, quite different from the bonding in a bulk material.
In a monowire, there are only two nearest neighbors, and therefore it is expected that the
bond length minimizing the total energy be smaller than in the bulk.
This is indeed the case, as can be seen in Table~\ref{tab:lattpar}, where calculated
bond lengths for monowires and bulk are listed, together with the  experimental bulk values.
Our bulk GGA calculations for the equilibrium bond lengths are in very close agreement
with the experimental values, and slightly underbonding. 
Our corresponding LDA calculations (not shown) yield, as expected, 
slightly shorter bond lengths, and overbond.
Our nonmagnetic nanowire calculations compare well with existing ones in 
Refs.~\onlinecite{sanchezportal1999,torres1999}, and \onlinecite {bahn2001}.
We should perhaps stress again that, strictly speaking, a tip-suspended 
wire will not have a quasi-stable configuration
at the bond length which minimizes the total energy, but at a slightly
larger value since it is rather the string tension than the total energy which should attain a local
minimum.\cite{tosatti2001_tension} Nevertheless, for simplicity, in the remainder of this paper, 
the bond length which minimizes the total energy will be called the
equilibrium bond length. 

Table~\ref{tab:lattpar} also shows our calculated mean-field magnetic moments at the equilibrium bond lengths.
The scalar relativistic calculations (SR) predict the Os and Ir wires to be magnetic at the equilibrium
bond length. In contrast, the fully relativistic calculations (FR) predict a much smaller moment for Os compared
to the SR calculation, 
no moment at all for Ir, and then, quite unexpectedly, a substantial moment in the Pt wire.
Thus, the spin-orbit coupling is seen to have a profound effect 
on the existence and magnitude of the magnetic moments. 

The rightmost column in Table~\ref{tab:lattpar} lists the experimental atomic ground
state configuration, showing that  the free Os, Ir, Pt and Au atoms have spin moments
4, 3, 2, and 1~$\mu_B$, respectively. Thus, the predicted wire moments are much smaller than the
magnetic moments of the free atom.

An interesting side question is whether there exists a substantial magnetostrictive effect in the wires, 
i.e., if the appearance of a magnetic moment in itself causes the equilibrium bond length to increase.
Although the calculated wire magnetic moments are quite large in some cases, we find that this
has almost no effect on the equilibrium bond length. The calculated equilibrium bond lengths for the magnetic wires are
indeed always larger, but only very slightly so, typically one or two hundredths of an {\AA}ngstr\"om.
In fact, the strictive effect of spin-orbit coupling is as large or larger (while still a small effect). 
For the Os and Au wires, the bond length decreases when the spin-orbit coupling is taken into account,
whereas in Ir and Pt it increases. De Maria and
Springborg\cite{demaria2000} also calculated 
a similar decrease of the Au monowire bond length.
 
\begin{table}
\caption{
\label{tab:energies}
Energy difference per atom between wire and bulk, and between the ferromagnetic and nonmagnetic wire.   
SR = scalar relativistic calculation, FR = fully relativistic calculation.
NM = nonmagnetic calculation, FM = ferromagnetic calculation.
}
\begin{ruledtabular}
\begin{tabular}{ccccc}
&  $E_{\rm wire} - E_{\rm bulk}$  &  $E_{\rm NM} - E_{\rm FM}$  &  $E_{\rm NM} - E_{\rm FM}$  \\
&              (eV)               &      (meV)                  &     (meV)                   \\
metal     &        FR                       &       SR                    &      FR                     \\
\hline
\hline
&                                 &                             &                             \\            
Os        &      5.5                        &    18                       &      6                      \\  
Ir        &      4.5                        &    43                       &      -                      \\  
Pt        &      4.0                        &     -                       &      8                      \\  
Au        &      2.3                        &     -                       &      -                      \\  
\end{tabular}

\end{ruledtabular}
\end{table}

In order to analyze the stability of wire formation as well as the stability of the magnetism in the wires, 
we calculated the energy gain when the wire is allowed to spin polarize,
and also the energy difference between wire and bulk. 
The results are displayed in Table~\ref{tab:energies}.
For Au, a bulk atom is around 2~eV more stable than the monowire, whereas for Os, Ir and Pt, this
energy difference is about twice as large. This rationalizes why wire formation is easiest in Au.
Energy differences between monowire and bulk have been reported earlier for Pt and Au, 
and our results are in good agreement with those 
calculations.\cite{sanchezportal1999,torres1999,bahn2001}
The energy gain due to spin polarization is of course a much smaller
quantity, and differs greatly from element to element. 
For example, in the scalar relativistic calculations, the energy gain for 
Ir is much greater than that in Os, although the moment is larger in Os than in Ir.
It is also very sensitive to the spin-orbit coupling. In the case of Os,
the relative stability of the magnetic solution  drops from 18 to 6~meV
when spin-orbit coupling is introduced. This drop for the Os wire is to a large
extent due to the magnetic moment being much smaller in the FR calculation. 
Such a small magnetic energy gains suggests that cryogenic temperatures 
could be required in order for the slow fluctuation
regime to be reached, and magnetism to be observable, in these nanowires.

\subsection{Magnetic moments}
\begin{figure}[h]
\psfig{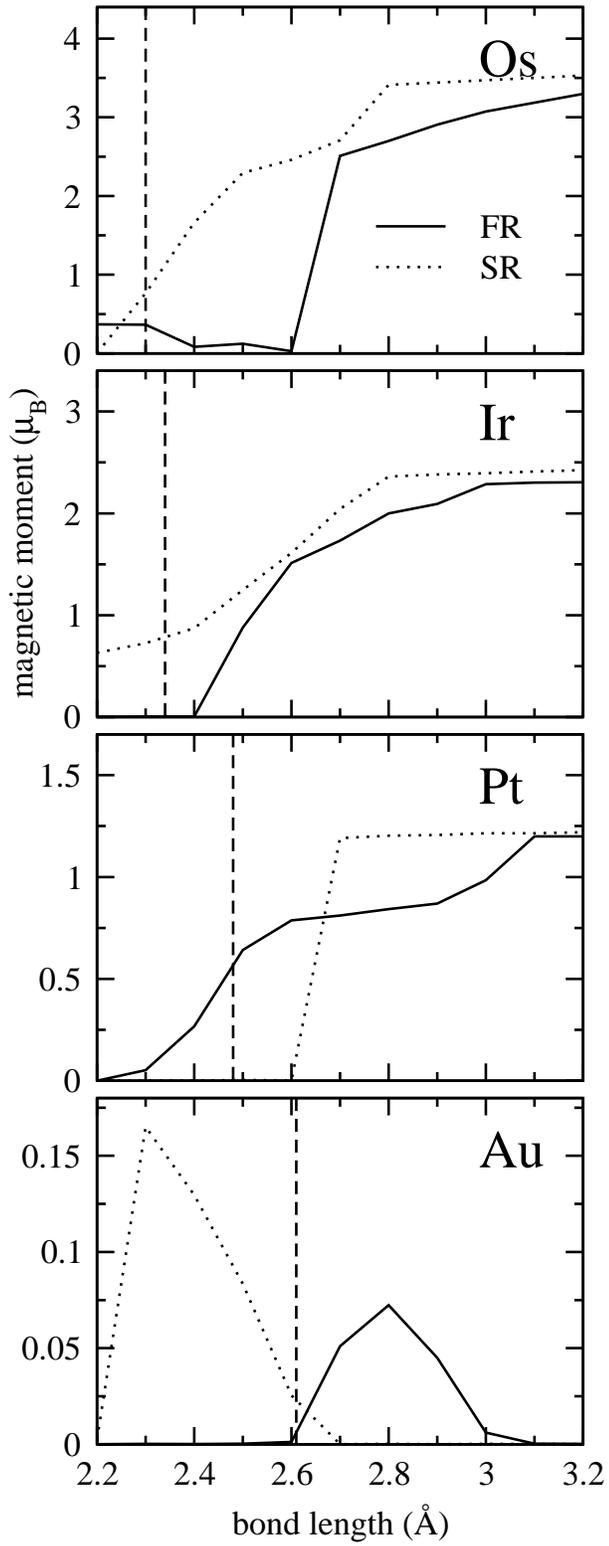}
\caption
{
Total magnetic moments, both with spin-orbit coupling (FR) and without (SR), per atom as a function of bond length.
The dashed vertical line points out the equilibrium bond length. 
\label{fig:total_magnetic_moment}
}
\end{figure}

The magnetic moments per atom monowire as a function of bond length are shown in Fig.~\ref{fig:total_magnetic_moment}.
The solid lines refer to the fully relativistic calculations (FR), and the dotted lines to the scalar relativistic (SR)
calculations.
The first thing to note is that all the metals studied exhibit a magnetic moment for  values of the 
bond lengths at or close to equilibrium. 
Ir and Au merely need a slight stretch in order to spin polarize.
Another general feature is that the magnetic profiles for the SR and FR calculations are very different. 

For instance, the SR calculation for
Pt predicts this metal to be magnetic only for stretched wires, whereas the FR calculation predicts it to
be magnetic in the whole range of bond lengths studied (2.2 {\AA} to 3.2 {\AA}).
Also the Os wire spin polarizes in the whole range of bond lengths studied. Unexpectedly, for this metal the FR calculation predicts
the magnetic moment to initially decrease with stretching whereas the SR calculation finds a monotonically increasing 
magnetic moment.

For  Os, Ir and Pt, the magnetic moment reaches a  plateau value for very large bond lengths 
(around and beyond 3.2 {\AA}, so large that the wires are most probably since long broken).
The value of this plateau magnetic moment is close to, even if still below, the atomic spin moment.
In Au, the situation is quite different from that of the other metals. 
The Au wire acquires a very small magnetic moment, less than 0.1~$\mu_B$
in the FR calculation, for slightly stretched bond lengths. 
With further stretching, the moment disappears again. 
Of course, it will eventually reappear at larger (but unphysical) bond lengths, because 
the free Au atom has a filled $5d$ shell and one unpaired
$6s$ electron, giving a pure $s$ moment of 1~$\mu_B$. 

In order to shed some light onto the mechanisms behind the magnetic profiles displayed in
Fig.~\ref{fig:total_magnetic_moment}, we will now analyze the electronic structure of the wires, using
band structures and the energy positions of $s$ and $d$ levels relative to
one another and to Fermi.

\subsubsection{Relative positions of $s$ and $d$ levels}
A determining factor for the magnetic state of transition metal atoms is the close competition between the 
$s$ and $d$ states.
According to the standard Aufbau principle of orbital filling,
the $(n+1)s$ orbitals should fill before the $nd$ orbitals, where $n$ is the principal atomic orbital quantum number. 
However, this rule is often broken for heavier elements.
The reason is that due to relativistic effects influencing the kinetic
energy of the orbitals
the relevant $s$ and $d$ levels are very close in
energy, so which one becomes populated in the end may depend
on a number of factors such as the fine balance between the repulsion of the other orbitals in the shell, 
the energy gained from completing a $d$ shell
(if possible), the energy cost associated with populating both orbitals in the $s$ shell, 
and the form of the orbitals (due to different $n$).

In bulk, on a surface, or in a wire, the situation is further 
complicated by hybridization and the accompanying broadening
of the atomic levels into bands. 
Magnetism may not even appear at all, since for broad enough bands, the
exchange energy gain due to spin polarization cannot match the
increased cost in terms of kinetic energy. 
This is the situation for the bulk $5d$ transition metals and also 
for wires with very short bond lengths.

\begin{figure}[h]
\psfig{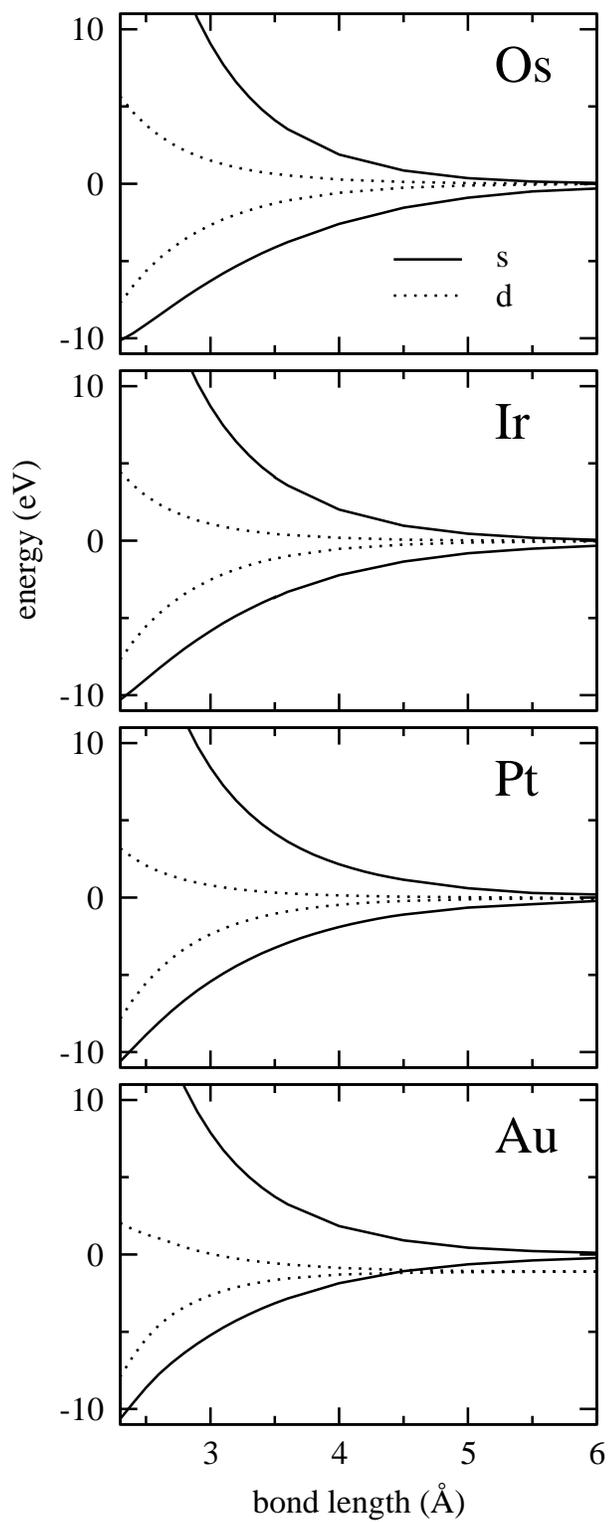}
\caption
{
Bottom and top of the $s$ and $d$ bands as a function of bond length. 
The Fermi energy is at zero. 
\label{fig:bottom_top}
}
\end{figure}

In order to quantify the relative positions of the $s$ and $d$ levels for our wires, we plotted the 
bottom and top of the $s$ and $d$ bands as a function of bond length, see Fig.~\ref{fig:bottom_top}. 
The bottom and top of a band have been estimated using the Wigner-Seitz rule, 
so that the top is taken as the energy where the wave function is zero,
and the bottom of the band is that where the derivative of the wave function is zero. 
This qualitative measure of the bandwidth 
tells us the relative positions of the $s$ and $d$ states, especially for large bond lengths where the 
bands narrow into atomic-like levels. 
Calculations must be taken up to very large bond lengths (6 {\AA}), 
in order to recover the situation close to that of free atoms.
As we will see, this analysis of the relative band positions catches the main trends for the
wire magnetic moments.
In Os and Ir, we see that the $d$ level is slightly above the $s$ level at the atomic limit, with the result that 
the $s$ shell will fill up, giving the atomic configurations $d^7s^2$ and $d^8s^2$, respectively.
This matches with the overall tendency of the magnetic moments in Os and Ir to increase as the wire is stretched (at least for
large enough bond lengths) in the following way. Two mechanism are at
work. The first one, valid as long as the band widths are still
substantial, is that as the $d$ band width decreases, 
the spin polarization within the $d$ shell increases due to exchange. 
The second mechanism, valid in the atomic limit, is that as the $s$ shell fills up, the
number of $d$ electrons decreases, which, equivalently, results in an increased magnetic moment.
In Pt, $s$ and $d$ levels are essentially degenerate, and consequently the
$s$ shell never fills up completely (atomic configuration $d^9s^1$).
In Au the $d$ level lies clearly beneath the $s$ level, and so the $d$ shell will be fully 
occupied for large enough bond lengths. This is the reason why the $d$ magnetism in the Au wire
disappears at larger bond lengths. 

\subsubsection{Band structures}
\begin{figure}[h]
\psfig{figure=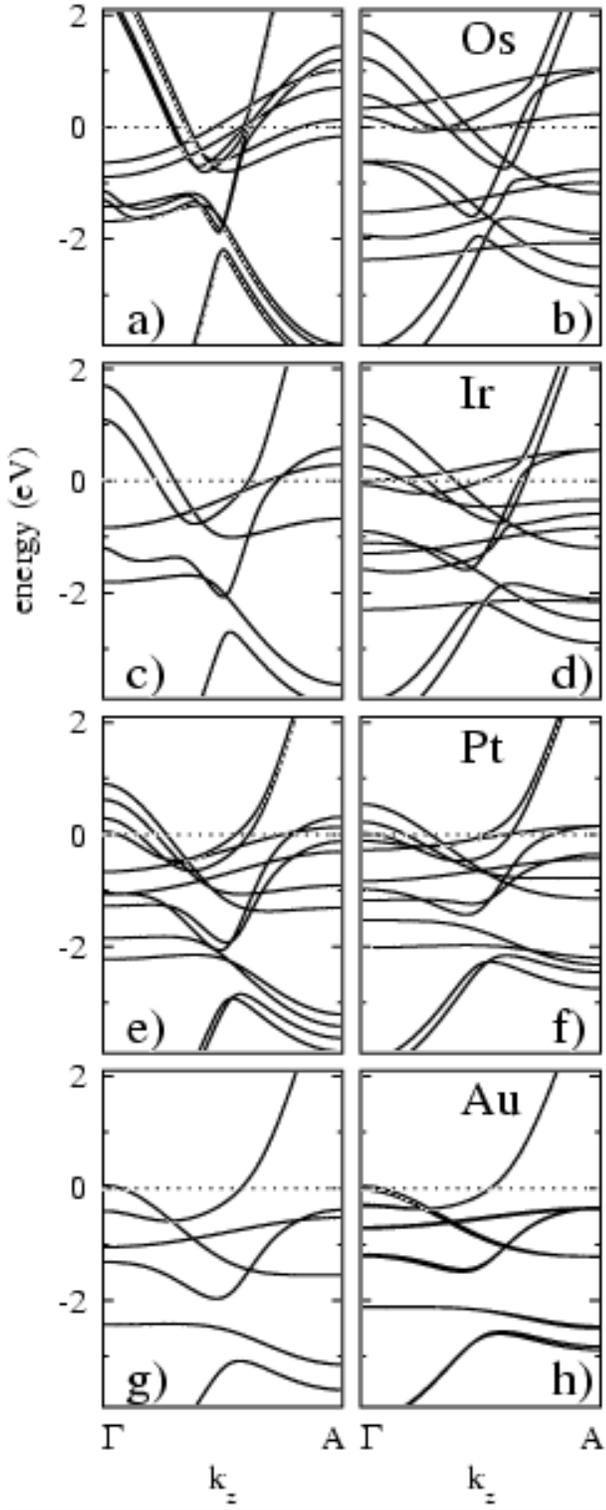,width=8.0cm}
\caption
{
Band structures, along the wire direction, at two different bond lengths 
(the equilibrium one, and a larger of 2.8 {\AA}) for each element.
The Fermi energy is at zero.
Band doubling (present in panels a, b, d, e, f, and h) indicates spin splitting due
to magnetic order.
\label{fig:band_structure}
}
\end{figure}

Some more detailed insight regarding the shape of the magnetic profiles can be gained by analyzing the
band structures, and how they change as a function of bond length.
Band structures for two different bond lengths, the equilibrium bond length, and a larger one of 2.8 {\AA}, 
roughly representing two magnetic regimes, are shown in Fig.~\ref{fig:band_structure} for each of our four elements.
The bands run from the zone center, $\Gamma$, to the zone edge, A, in the direction of the wire.

\begin{figure}[h]
\psfig{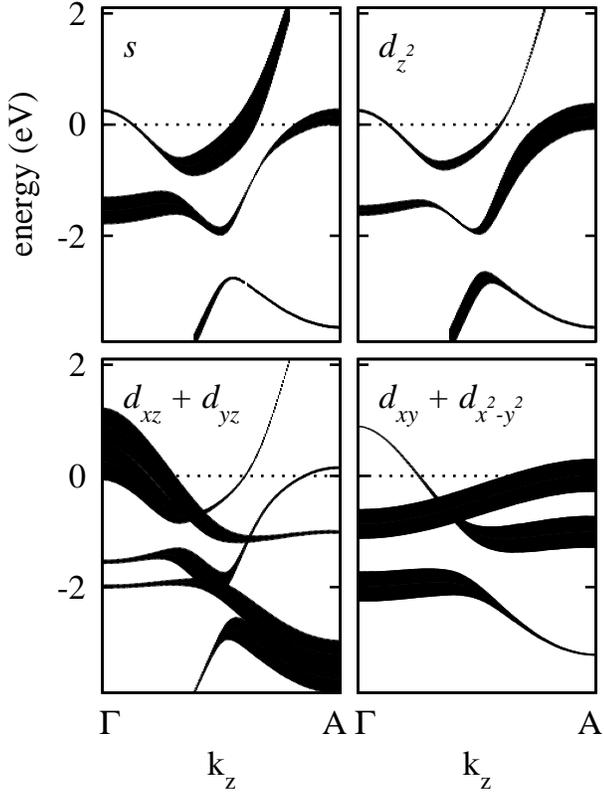}
\caption
{
Character-resolved band structure for nonspinpolarized Pt, along the wire direction.
The Fermi energy is at zero.
\label{fig:fatbands}
}
\end{figure}

The character of the bands close to the Fermi level is of critical importance for the moment formation,
and therefore we also show character-resolved bands, see Fig.~\ref{fig:fatbands}. 
We found it useful to split up the $d$ character into $d_z$, $d_{xz}+d_{yz}$ and $d_{xy} + d_{x^2-y^2}$, and so,
Fig~\ref{fig:fatbands} has four panels, displaying separately the 
$s$, $d_z$, $d_{xz}+d_{yz}$ and $d_{xy} + d_{x^2-y^2}$ characters
of the bands.
The vertical error bars, or ``thickness'', of the bands indicate the relative character weight. 
The data in Fig~\ref{fig:fatbands} has been taken from a calculation for Pt.
However, for the other metals, the relative weight of the orbitals for each band is qualitatively similar to the 
one shown. 
From Fig.~\ref{fig:fatbands}, we see that almost all bands in the vicinity of the Fermi level have predominantly $d$ character. 
In fact, there are only two bands with some $s$ character crossing the Fermi level (see upper left
panel in Fig.~\ref{fig:fatbands}). Of these, the highest lying band crosses the Fermi level halfway between the 
zone center and zone edge. This band is almost purely $s$ at that crossing. 
For Ir, Os and Pt, this band actually crosses the Fermi level twice. However, the degree of $s$
character for this band diminishes rapidly as the reciprocal lattice vector approaches the zone center $\Gamma$, 
i.e., the second crossing is $d$-dominated, as is evident from Fig.~\ref{fig:fatbands}.
The second one of the two $s$-containing bands crosses the Fermi level close to the zone edge (A).
At that point, it has some $s$ character, but is in fact dominated by $d_z$ character.

At  $\Gamma$ and A, both of them critical points by symmetry, 
all band dispersions are horizontal, giving rise to very sharp band edge
van Hove singularities, a feature due to the one-dimensionality of the systems. 
Since the bands have mostly $d$ character at the edges, the exchange energy
gain will be rather large if a band spin-splits so that one of 
the spin-channel band edges ends up above the Fermi level, and the other one below.
Strictly speaking, the spin-orbit coupling will mix the two spin channels so that, in general, an eigenvalue
will have both majority and minority spin character. 
However, in the present calculations, this mixing is so small, typically just a few percent, that it is
irrelevant for the qualitative discussion we make here.
Thus, if a band edge ends up sufficiently near the Fermi level, we may expect a magnetic moment
to develop. While apparently similar to the magnetization of the jellium
wire, magnetism here is much more substantial, the $d$ states involving
a much stronger Hund's rule exchange.  
We now go through all four metals, starting with Os, analyzing how the band edges move as a function of bond length,
and how this affects the magnetic state of the wires.

{\it Os:}
The magnetism in the Os wire has two regimes, one for bond lengths below 2.6 {\AA}, and one for bond lengths above this value. 
Below 2.6 {\AA}, the magnetic moment actually decreases with increasing bond length.
At the equilibrium bond length, only one band edge (of mostly $d_z$ character and some $s$ character), 
at A, has spin-split around the Fermi level (see panel a in Fig~\ref{fig:band_structure}).
This gives rise to a small moment of a few tenths of a $\mu_B$.
As the bond length increases, this band edge moves downward, through the Fermi level, 
and the magnetic moment is killed off.
At the same time, the band edges (at $\Gamma$ and A) of the rather flat $d_{xy} + d_{x^2-y^2}$ 
band come sufficiently close to the Fermi level, causing a large splitting
(see panel b). This results in a rapid increase of the magnetic moment, creating the second, large-moment, magnetic regime.

{\it Ir:}
With one more electron than Os, the bands of the Ir wire lie generally
deeper. At the equilibrium bond length, the band edge responsible for the
low-moment regime for Os lies well below the Fermi level and is
inactive. With increased bond length, the A edge 
of the flat $d_{xy} + d_{x^2-y^2}$ band gradually sinks 
toward the Fermi level and eventually causes a large magnetic
splitting. Thus, the whole magnetic regime in Ir is similar to 
the large-moment regime in Os.

{\it Pt:}
In Pt, the very same flat $d_{xy} + d_{x^2-y^2}$ band leading to Hund's
rule magnetism in Os and Ir behaves here in the opposite way. At very small bond lengths (2.2 {\AA}), 
this band is entirely occupied, and moves upwards (instead of downwards) with increased bond lengths.
As the edge at A touches the Fermi level, a magnetic moment develops. Two other bands, 
a $d_{z^2}$-dominated one with band edge at A and a $d_{xz}+d_{yz}$-dominated one
with band edge at $\Gamma$ are also important. They are just slightly higher in energy than the first band edge,
and with increasing bond length, they move to lower energies. Thus, these three band edges become increasingly
degenerate with stretching, and split around the Fermi level at 2.4 {\AA}, causing a rapid increase in the magnetic moment.

{\it Au:}
For Au, the $d$ bands causing the magnetism  in Os, Ir, and Pt lie well
below the Fermi level and cannot give rise to a magnetic moment.
The magnetically active band edge is at  $\Gamma$, and belongs to a band
with relatively high dispersion and $d_{xz}+d_{yz}$ character.
With increasing bond length, this band edge moves downward, and as it passes through the Fermi level
it creates a small magnetic moment. As can be seen in Fig.~\ref{fig:band_structure},  panel h, the spin splitting of the
band edge is really very tiny, and the magnetism in Au is reminiscent of the magnetism of the jellium cylinder, i.e.,
a band-edge phenomenon rather than Hund's rule driven spin polarization.
Further stretching causes this edge to sink below the Fermi energy, and the magnetic moment 
consequently disappears. It is not clear at present whether this 
moment may be of any real physical significance. 

\subsection{Ballistic conductance channels}

As seen from the above discussion of the nanowire band structures,
spin-splitting of bands does alter the number of bands --- or channels --- $n$
crossing the Fermi level.  
By virtue of the Landauer formula 
\begin{equation}
G = \frac{e^2}{h}\sum_i \tau_i,
\end{equation}
where $\tau_i$ is the transmission through channel $i$,
the ballistic conductance measured has, in units of $\frac{1}{2}G_0 = e^2/h$, precisely the
number of bands $n$ crossing the Fermi level as its upper limit.
Thus, the conductance through the wires should be affected by magnetism.

\begin{figure}[h]
\psfig{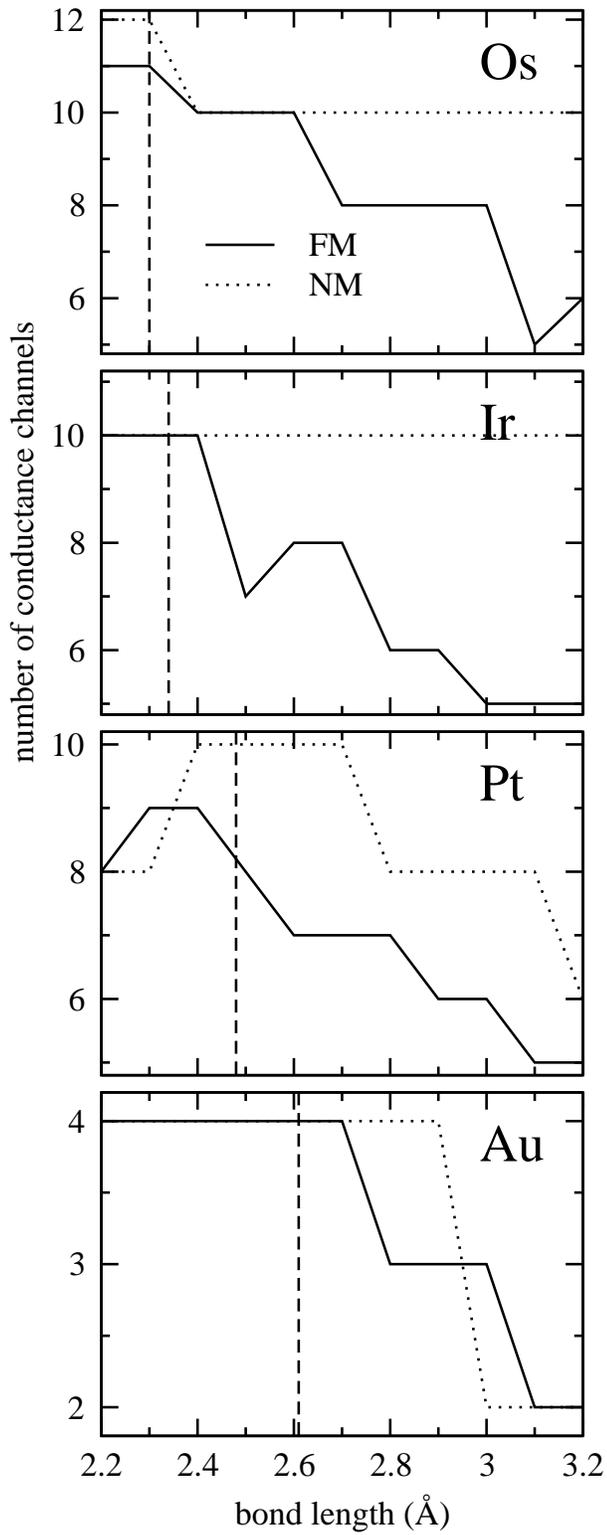}
\caption
{
The number of open conductance channels as a function of bond length.
FM = ferromagnetic calculation; NM = nonmagnetic calculation.
The dashed vertical line points out the equilibrium bond length. 
\label{fig:conductance_channels}
}
\end{figure}

Fig.~\ref{fig:conductance_channels} shows how
the number $n$ of conducting channels is influenced by nanowire spin-polarization and bond length.
For Os and Pt in their magnetized state at the equilibrium bond length, 
$n$ is large, 11 and 8, respectively, against 12 and 10 in the nonmagnetic state.
Magnetism has decreased the number of channels, but not dramatically so.
Should all these channels transmit fully, a large ballistic conductance of
$4 G_0$ for Pt or $5.5 G_0$ for Os would ensue, to be compared with
nonmagnetic values of $5 G_0$ and $6 G_0$, respectively. 

In reality however most of the open channels have $d$ character. While the
conductance of the broad band $s$ channels is generally close to one owing to nearly
complete transmission, that of the narrow band $d$ channels is much
smaller, with a high reflection at the lead-wire junction, 
generally dependent on the detailed junction geometry. 
Of the conductance channels in these metals, two have $s$ character, both in the spin-polarized and 
nonmagnetic calculations, bringing an expected contribution close to $G_0$ to the total conductance.
All the other channels have predominantly $d$ character. Their contribution to the conductance
is therefore expected to be much smaller than $\frac{1}{2} G_0$ per channel.
We may thus expect these wires to have a conductance above $G_0$ but well 
below $4 G_0$ and $5.5 G_0$, respectively.
Since the scattering of the $d$ waves at the junctions depends highly 
on the geometry, whose details will change at every realization, 
we also expect the conductance histograms to exhibit peaks 
that could be both broad and poorly reproducible. 
For Ir, our calculations indicate that the conductance at the equilibrium
bond length should lie between $G_0$ and $5 G_0$.
For these three metals, according to our calculations the number of conductance
channels decreases --- by and large --- as the wire is stretched.
However, the disappearing channels are always $d$-dominated.

Of the metals Os, Ir and Pt, measured conductance histograms have been published only for Pt so far.
Smit {\it et al.}\cite{smit2002} find a large, broad peak 
centered around $1.5 G_0$ and a smaller bump centered around $2.2 G_0$. 
The conductance histograms reported by Yanson\cite{yanson_thesis} 
are similar in structure, but the positions are shifted, to around $1.7 G_0$ and $3 G_0$.
Rodrigues {\it et al.}\cite{rodrigues} find a peak centered around $1.4 G_0$, and in addition
a peak at very low conductance, around  $0.5 G_0$.

In the Au wire, we find theoretically four open conductance channels.
Two of these are $s$ dominated, just as
for the other metals, and two are $d$ dominated. However, the $d$ channels
are merely touching the Fermi level, and are therefore expected to have a very marginal effect on the conductance.
Experimentally, gold nanowires yield a rather sharp peak between $0.9 G_0$
and $G_0$, confirming that the $d$ influence is probably very small.

\section{Conclusions}
In conclusion, our calculations suggest that the Os, Ir, and Pt monatomic
nanowires should exhibit spontaneous Hund's rule superparamagnetism for values of the
bond length at equilibrium or --- in the case of Ir --- slightly above.
The energy gain connected with the magnetic state is small, less than 10~meV
for Os and Pt at the equilibrium bond length.  Au nanowires also
theoretically magnetize, but the calculated energy gain is an order of magnitude smaller than
for the other metals.
From a methodological point of view, 
the spin-orbit coupling is found to be crucial for a correct description of the magnetic state,
as is probably the use of all-electron techniques.\cite{bahn2001}

How might this magnetism be detected experimentally? 
Merely measuring the conduction through the wire 
at one single temperture and magnetic field strength will most probably not give
conclusive information regarding the magnetic state of the atoms in the wire, since
the tranmission through $d$ channels is rather poor and 
vary greatly with geometry and can hardy be regarded as quantized.
A key experiment would be to measure ballistic conductance as a function of 
temperature and of external magnetic field. At high temperature and zero
field the nanowire should be nonmagnetic, due to fast fluctuations.
High field and low temperature would take the nanowire to a magnetic, or in any case to a 
slowly fluctuating superparamagnetic regime. In this transition the number of conductance channels
should diminish, and so should the conductance --- even if not by very much.
At sufficiently low temperatures, the conductance should definitely be field sensitive.
Such a behavior would be a clear indication of a superparamagnetic state.
 
In some situations, more majority bands may cross the Fermi level than do minority bands, 
leading to partial spin-polarization of the transmitted electron
current. If this current could be measured, it would be a very direct way of 
confirming the existence of a superparamagnetic state.

Fractional conductance peaks have been observed experimentally, for example the  $\frac{1}{2}G_0$ peak reported
by Ono for Ni,\cite{ono1999} and very recently by Rodrigues {\it et al.} for Co, Pd and Pt,\cite{rodrigues}
at room temperature and zero field. 
These results are intriguing, since we expect that the $s$ channel 
alone should yield a conductance larger than that.
The peaks observed in Co, Pd, and Pt, centered around  $\frac{1}{2}G_0$, are rather broad, which 
suggests that they might not be caused by one single fully transmitting spin-polarized channel, but perhaps by
several poorly conducting channels.
We discussed in previous work,\cite{smogunov1} a
possibility to obtain conductance $G_0$ from a magnetic transition metal
nanowire with a magnetization reversal occurring inside the nanowire. 
This could further drop to $\frac{1}{2}G_0$ in an asymmetrical situation, with a net prevalence of majority spins
over minority spins. 
Although it is not inconceivable that this might occur in Co and Ni,
we are unable to explain at the moment how that kind of state could be
sustained in Pt, and by extension in Pd too, at the experimental conditions
of zero field and room temperature. It would anyway be interesting to see
the effect of cooling and of an external field on these results.

\acknowledgments
A.D. acknowledges financial support from 
the European Commission through contract no. HPMF-CT-2000-00827 Marie Curie fellowship,
STINT (The Swedish Foundation for International Cooperation in Research and Higher Education),
and NFR (Naturvetenskapliga forskingsr{\aa}det).
Work at SISSA was also sponsored through TMR FULPROP, MUIR (COFIN and FIRB) and by INFM/F.
Ruben Weht is acknowledged for discussions, and for double-checking some of the calculations using the WIEN97 code.
J. M. Wills is acknowledged for letting us use his FP-LMTO code.
We are also grateful to D. Ugarte for sharing with us the unpublished results of Ref.~\onlinecite{rodrigues}.

%%%%%%%%%%%%%%% REFERENCES 

%
%%%%%%%%%%%%%%% FIGURES
 % MAGNETIC MOMENTS
%
%BOTTOM_TOP
%
% BAND STRUCTURES all elements
%
% FAT BANDS 
%
% CHANNELS

\begin{thebibliography}{99}

\bibitem{kondo2000_helical}
K. Kondo and K. Takayanagi,
Science {\bf 289}, 606 (2000).

\bibitem{rodrigues2000}
V. Rodrigues, T. Fuhrer, and D. Ugarte, \prl {\bf 85}, 4124 (2000).

\bibitem{smit2001}
R. H. M. Smit {\it et al.}, 
\prl {\bf 87}, 266102 (2001)

\bibitem{wees1988}
B. J. van Wees {\it et al.},
Phys. Rev. Lett. {\bf 60}, 848 (1988).

\bibitem{gulseren1998}
O. G\"ulseren, F. Ercolessi, and E. Tosatti,
Phys. Rev. Lett. {\bf 80}, 3775-3778 (1998).

\bibitem{tosatti2001_tension}
E. Tosatti {\it et al.}, 
Science {\bf 291}, 288 (2001).

\bibitem{ohnishi1998}
H. Ohnishi, Y. Kondo, and K. Takayanagi,
Nature, {\bf 395}, 780 (1998).

\bibitem{yanson1998}
A. I. Yanson, {\it et al.},
Nature {\bf 395}, 783 (1998).

\bibitem{gambardella2002}
P. Gambardella {\it et al.}, 
Nature {\bf 416}, 301 (2002).

\bibitem{heehong2001}
B. H. Hong {\it et al.}, 
Science {\bf 294}, 348 (2001).

\bibitem{zabala1998}
N. Zabala, M. J. Puska, and R. M. Nieminen,
Phys. Rev. Lett. {\bf 80}, 3336 (1998); 
Comment and Reply, {\it ibid.} {\bf 82} 3000 (1999).
	 
\bibitem{note1} In the case of Pt, we also performed antiferromagnetic calculations, but
found this magnetic configuration to be unstable
with respect to ferromagnetic ordering.

\bibitem{blugel1992}
S. Bl\"ugel,
Phys. Rev. Lett. {\bf 68} 851 (1992).

\bibitem{redinger1995}
J. Redinger, S. Bl\"ugel, and R. Podloucky,
Phys. Rev. B {\bf 51}, 13852 (1995).

\bibitem{sanchezportal1999}
D. Sanchez-Portal, {\it et al.},
Phys. Rev. Lett. {\bf 83}, 3884-3887 (1999).

\bibitem{torres1999}
J. A. Torres {\it et al.},
Surf. Science {\bf 426}, L441-L446 (1999).


\bibitem{dft}
P. Hohenberg and W. Kohn, Phys. Rev. {\bf 136},  B864  (1964);
W. Kohn and L. J. Sham, Phys. Rev. {\bf 140},  A1133  (1965).

\bibitem{wills}
J. M. Wills, O. Eriksson, M. Alouani, and O. L. Price, 
in {\it Electronic Structure and Physical Properties of Solids},
edited by H. Dreyss\'e (Springer, Berlin, 2000).

\bibitem{gga}
J. P. Perdew, in {\it Electronic Structure of Solids 1991},
edited by P. Ziesche and H. Eschrig (Akademie Verlag, Berlin, 1991);
J. P. Perdew, K. Burke, and M. Ernzerhof,
\prl {\bf 77}, 3865 (1996);
J. P. Perdew, K. Burke, and M. Ernzerhof,
\prl {\bf 78}, 1396 (1997);
Y. Zhang and W. Yang,
\prl {\bf 80}, 890 (1998);
J. P. Perdew, K. Burke, and M. Ernzerhof,
\prl {\bf 80}, 891 (1998).

\bibitem{lda}
D. M. Ceperley and B. J. Alder,
\prl {\bf 45}, 566 (1980);
J. P. Perdew and A. Zunger,
\prb {\bf 23}, 5048 (1981).

\bibitem{wien97}
P. Blaha, K. Schwarz, and J. Luitz, 
computer code WIEN97 (Vienna University of Technology, Vienna, 1997), 
[Improved and updated UNIX version of the original copyrighted WIEN code, which was published by
P. Blaha, K. Schwarz, P. Sorantin, and S. B. Trickey, Comput. Phys. Commun. {\bf 59}, 339 (1990)].

\bibitem{bahn2001}
S. R. Bahn and K. W. Jacobsen,
\prl {\bf 87}, 266101 (2001).

\bibitem{demaria2000}
L. De Maria and M. Springborg,
Chem. Phys. Lett. {\bf 323}, 293 (2000).

\bibitem{smit2002}
R. H. M. Smit {\it et al.}, 
Nature {\bf 409}, 906 (2002).

\bibitem{yanson_thesis}
A. I. Yanson, Thesis, Universiteit Leiden, The Netherlands (2001). 

\bibitem{rodrigues}
V. Rodrigues, J. Bettini, P. C. Silva, and D. Ugarte, preprint.

\bibitem{ono1999}
T. Ono, Y. Ooka, H. Miyajima, and Y. Otani,
Appl. Phys. Lett. {\bf 75}, 1622 (1999).

\bibitem{smogunov1}
A. Smogunov, A. Dal Corso, and E. Tosatti,
Surf. Science {\bf 507}, 609 (2002). 

\end{thebibliography}
 \end{document}